\documentclass[12pt]{article}
\usepackage{graphicx}
\usepackage{mathptmx}
\usepackage[intlimits,centertags]{amsmath}
\usepackage{amssymb,amsfonts}
\usepackage[pdftex]{hyperref}
\usepackage{aas_macros}
\usepackage{enumerate}
\usepackage[symbol]{footmisc}
\usepackage{xfrac}
\usepackage[table,xcdraw]{xcolor}
\usepackage{sectsty}
\usepackage[normalem]{ulem}
\usepackage{setspace}
\usepackage{ragged2e}
\usepackage{combelow} 
\usepackage{etoolbox} 
\usepackage{cite}
\usepackage{fdsymbol}
\usepackage{pdfpages}
\usepackage[T1]{fontenc}
\usepackage{times}
\usepackage[compact]{titlesec}
\usepackage{geometry}
\usepackage[utf8]{inputenc}
\usepackage{enumitem,amssymb}
\usepackage{ragged2e}
\usepackage{wrapfig}
\newlist{thematic}{itemize}{8}
\setlist[thematic]{label=$\square$}
\usepackage{pifont}

\definecolor{darkgreen}{rgb}{0,0.5,0}
\definecolor{vermillion}{rgb}{0.89, 0.259, 0.204}

\newif\ifastrophysical
\astrophysicaltrue

\newif\iffundamental
\fundamentalfalse

\definecolor{header_color}{HTML}{741a07}
\subsectionfont{\color{header_color}}

\begin{document}
\begin{titlepage}
\includepdf{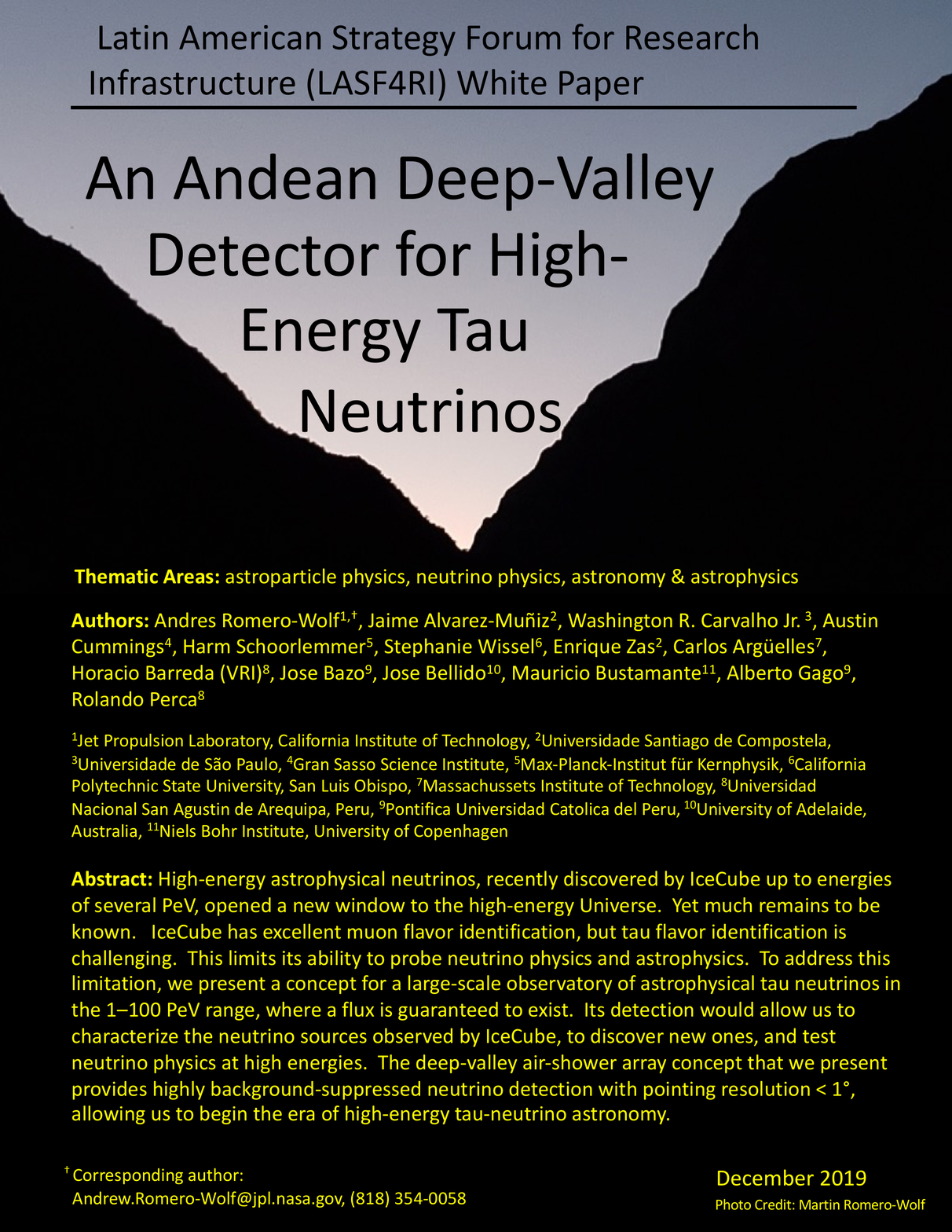}
\end{titlepage}

\pagenumbering{roman}

\normalsize
\pagebreak

\pagenumbering{arabic}

\subsection*{Scientific Context}

The last decade has seen the remarkable discovery of an astrophysical flux of high-energy ($>$10~TeV) neutrinos~\cite{Aartsen_2013} and subsequent characterization of the energy spectrum with observations up to several PeV~\cite{Aartsen_2014, Kopper_2016, Aartsen_2016, Haack_2017, Kopper_2018}. 
The recent observations of neutrinos from blazar TXS 0506+056 coincident with $\gamma$-rays provide the first evidence of an extragalactic neutrino source~\cite{Aartsen_2018, Kanata_2018}. 
As articulated in the Astro 2020 science white paper ``Astrophysics Uniquely Enabled by Observations of High-Energy Cosmic Neutrinos"~\cite{Vieregg_2019}, neutrinos carry unique information about the most energetic non-thermal sources in the universe. The concept presented here is designed to extend observations of neutrinos to higher energies, covering the range  of 1-100 PeV, with tau neutrinos ($\nu_\tau$) in response to the strategy advocated by the neutrino astrophysics community~\cite{Vieregg_2019} of a multi-observatory approach that would extend the science reach of neutrino observatories.

To date high-energy $\nu_\tau$ are the least studied of all neutrino flavors~\cite{Parke:2015goa}. Even though high-energy astrophysical neutrinos have been measured in the northern sky with muons~\cite{Aartsen:2018ywr} and for the full sky with all-flavor measurements~\cite{IC_2013,Aartsen_2014,Schneider:2019ayi}, isolating the astrophysical $\nu_\tau$ component has been challenging, with the first candidate events being identified after approximately ten years of data~\cite{Stachurska_ICRC_2019, Xu_ICRC_2019}. The concept presented here aims at the unambiguous measurement of the $\nu_\tau$ flux to enable characterization of the astrophysical sources. The sources are expected to produce primarily electron ($\nu_e$) and muon ($\nu_\mu$) neutrinos with the $\nu_\tau$ component resulting only from neutrino flavor conversion during propagation over astronomical distances. Characterization of the $\nu_\tau$ component of the astrophysical neutrino flux, in combination with the observations of IceCube and other detectors such as KM3NeT, are key observables to constrain the neutrino production mechanisms and physical conditions at the sources~\cite{Vieregg_2019, Anchordoqui_2005, Ackerman_2019} as well as new neutrino physics~\cite{Arguelles:2015dca,Bustamante:2015waa,Rasmussen:2017ert,Arguelles:2019rbn,Arguelles:2019tum}.


Neutrinos are an important part of multi-messenger studies since they are undeflected by magnetic fields and their attenuation length in the intergalactic medium is greater than the size of the universe. Extending the energy reach to 100~PeV with more than an order of magnitude increase in sensitivity would provide key information about the sources of high-energy neutrinos and energetic non-thermal sources in general~\cite{Vieregg_2019}. Constraining the neutrino flux from transient events such as stellar explosions, compact object mergers, and relativistic jets elucidates both the astrophysical particle acceleration mechanisms, as well as the fundamental physics of their interactions~\cite{Ackerman_2019}.

\begin{figure}[!t]
	\centering
		\includegraphics[width=0.7\linewidth,trim={3.0cm 4.0cm 5.0cm 4.0cm},clip]{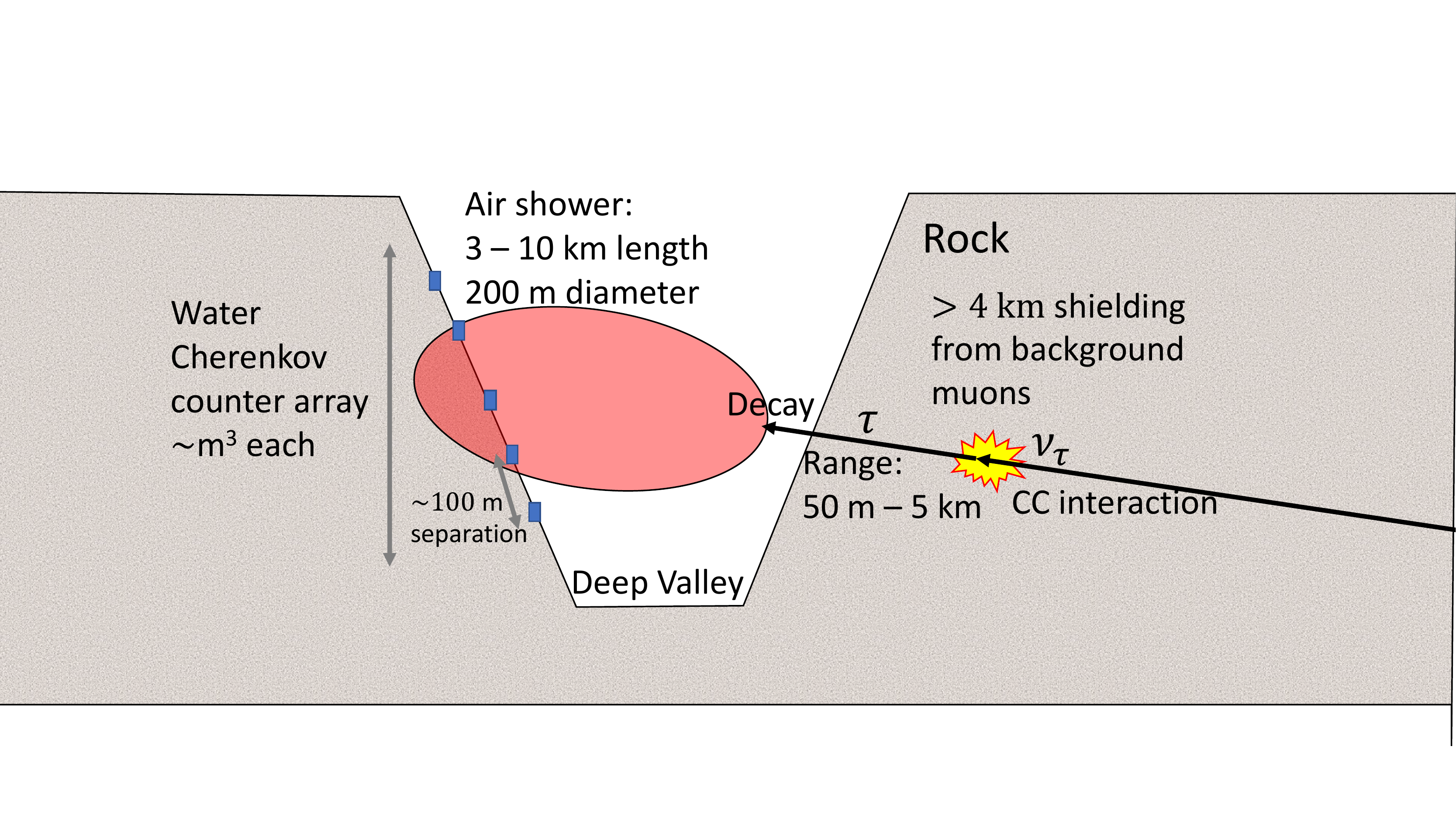}
	\caption{Detector concept schematic. A $\nu_\tau$ propagates through rock in the mountain. If a charged-current (CC) interaction occurs, a $\tau$ is produced with a range of 50~m--5~km (depending on the energy) before it decays. If the $\tau$ exits into the atmosphere it decays to produce an air shower. The electromagnetic component of the air shower extends $\sim\,$3-10~km (for the atmospheric density assumed in the fiducial site in this study), with a diameter of $\sim\,$200~m. Such an air shower is detectable by an array of water-Cherenkov tanks, with $\sim\,$m$^3$ volume each, separated by $\sim\,$100~m along and across the mountain slope.  }
	\label{FIG:concept}
\end{figure}

We present a concept for a deep-valley $\nu_\tau$ detector, an idea originally proposed by Fargion in 1999~\cite{Fargion_1999}, that could be implemented in the Colca Valley of Peru, called TAMBO (Tau Air Shower Mountain-Based Observatory), designed for $\nu_\tau$ detection in the 1-100~PeV band. The concept of detection is illustrated in Figure~\ref{FIG:concept}. Neutrinos propagating in rock have an interaction length of thousands of kilometers for 1-100~PeV energies~\cite{Ghandi_1998}. A $\nu_\tau$ can interact in rock to produce a $\tau$, transferring $\sim\,$80\% of its energy. The $\tau$ has a range of 50 m - 5 km for the corresponding energy range of 1-100 PeV before it decays. If the interaction occurs within this distance from a deep valley, the $\tau$ is likely to exit the rock into the air. Within this range, the $\tau$ may decay in air. Roughly 50\% of the $\tau$ energy goes into particles that produce extensive air showers. The longitudinal profile of the electromagnetic component of these air showers has a characteristic length of $\sim\,$10~km and a diameter of $\sim\,$200~m near the shower maximum. A deep valley with mountain separations comparable to this length scale and an array of small water-Cherenkov detectors, each of approximately one cubic-meter volume, separated by $\sim\,$100~m on the mountain slope, would be highly sensitive to these $\tau$-initiated air showers. The deep-valley topography also provides a significant increase to the geometric acceptance compared to a flat ground array sensitive to Earth-skimming air showers, as used in on-going experiments~\cite{Pedreira:2019}. The Colca Valley in Peru has been identified as an attractive site to implement this concept.

\subsection*{Objectives}


The science objectives of TAMBO are summarized in the table below.
\begin{figure}[!h]
	\centering
		\includegraphics[width=0.9\linewidth]{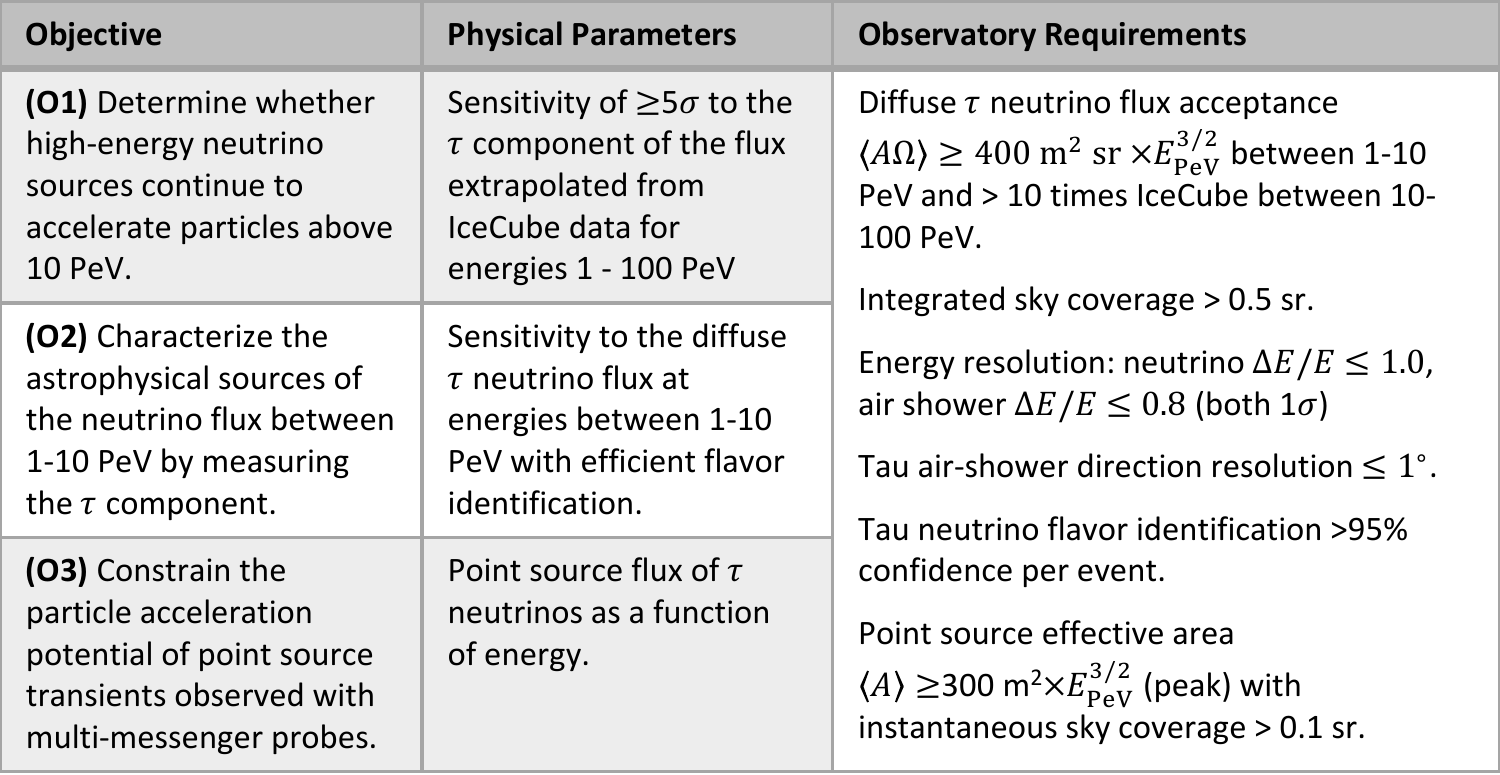}
	\label{TBL:reqs}
\vspace{-10pt}
\end{figure}

\begin{figure}
\centering
  		\includegraphics[width=0.5\linewidth]{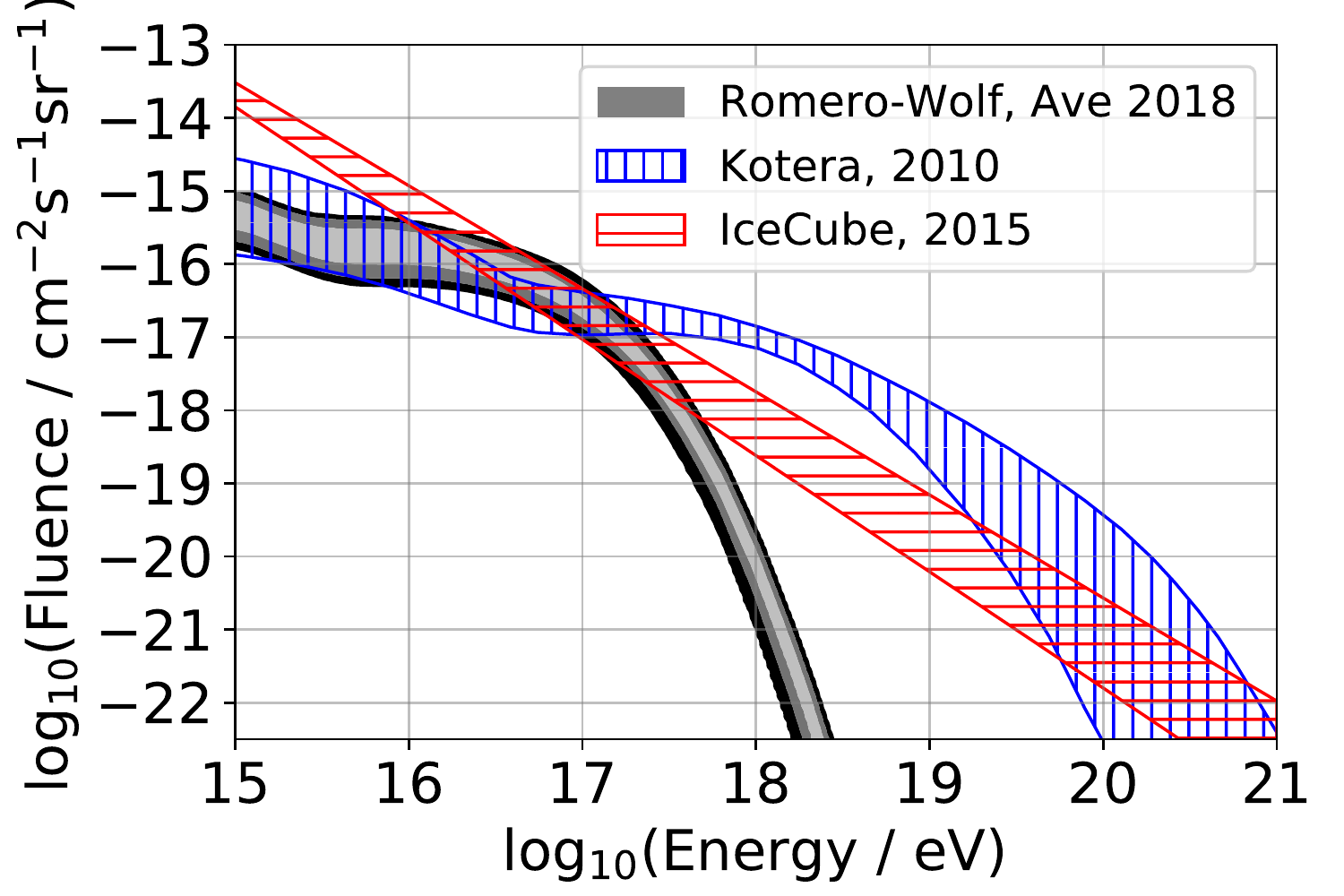}
		\caption{Neutrino fluxes that would be probed with TAMBOS. The band labeled IceCube, 2015 is the all-flavor IceCube neutrino flux ($\propto E^{-2.5}$)~\cite{IceCube_2015}, extrapolated to high energies. The Kotera, 2010~\cite{Kotera_2010} band shows a range of models that generally assumes higher energy cutoffs than the evidence provided by analysis of the Pierre Auger Observatory data~\cite{Auger_fits}. The band labeled Romero-Wolf \& Ave, 2018~\cite{Romero-Wolf_Ave_2018} is an example of the range of models that results from including constraints from Auger data. See \cite{Alves_Batista_2019, Heinze_2019} for similar results. This figure is adapted from \cite{Romero-Wolf_Ave_2018}.}
	\label{FIG:flux}
\end{figure}

Objective O1 is motivated by the question of whether the sources of high-energy neutrinos have a cutoff at $\sim\,$6~PeV, as suggested in some recent studies~\cite{Haack_2017}, and the detection of the cosmogenic neutrino flux, which is expected to dominate near 100 PeV. In Figure~\ref{FIG:flux}, we show the extrapolated IceCube flux compared to models of cosmogenic neutrino fluxes. The Kotera 2010~\cite{Kotera_2010} cosmogenic neutrino flux models assume cosmic-ray compositions and source energy cutoffs generally more optimistic than fits to the Pierre Auger Observatory data~\cite{Auger_fits}, while Romero-Wolf \& Ave 2018~\cite{Romero-Wolf_Ave_2018} use constraints from Auger data (see \cite{Alves_Batista_2019, Heinze_2019} for similar results). These models represent two extremes of what the cosmogenic neutrino flux could be and, in both cases, the flux is comparable to the extrapolated IceCube flux for energies approaching 100 PeV. 

Objective O2 is motivated by the lack of unambiguous detections of high-energy $\nu_\tau$ events to date. Observations in the 1-10 PeV energy band would be complementary to IceCube in that TAMBO is sensitive primarily to $\nu_\tau$ particles. IceCube is roughly equally sensitive to all neutrino flavors with good flavor identification sensitivity for $\nu_\mu$. Measurement of the diffuse $\nu_\tau$ flux would allow us to better understand high-energy neutrino production mechanisms~\cite{Bustamante_2019, Vieregg_2019} and fundamental physics with neutrinos ~\cite{Ackerman_2019,Arguelles:2015dca,Bustamante:2015waa,Rasmussen:2017ert,Arguelles:2019rbn,Arguelles:2019tum}.

Objective O3 is directly motivated by multi-messenger observations~\cite{Kanata_2018, Aartsen_2018}. The TAMBO sensitivity is focused in a region of $\sim\,$30$^\circ$ elevation angle and $120{^\circ}$ azimuth (see next section), which is complementary to IceCube's nearly all-sky instantaneous field of view in that it would enable a higher point source sensitivity. The region of the sky covered depends on the latitude of the observatory, which can in principle be placed anywhere there is a deep valley of appropriate size. Observations of $\nu_\tau$ events from point source transients such as stellar explosions, $\gamma$-ray bursts, AGN flares, and gravitational wave events would provide valuable information on the mechanisms underlying these events.

In addition to the objectives stated above, we are investigating additional science goals that could be achieved with TAMBO.  The unique topography could be used to improve our understanding of air-shower physics and to improve the measurement of the depth of shower maximum with water-Cherenkov detectors by detecting vertical showers and measuring the shower size at different slant depths. We are looking into the possibility of discriminating photons from hadrons and contribute to gamma-ray observations at the PeV region and beyond. It may also be possible to characterize the cosmic-ray anisotropy at PeV energies with new measurements from a latitude corresponding to the proposed observatory site in the Colca Valley, Peru. This would overlap with the HAWC and IceCube sky exposures~\cite{HAWC_IC_2017}, providing complementary information to characterize the anisotropy and further constrain its origin. 

\subsection*{Methodology}

The technical overview covers the observatory requirements listed in the table in page 2 and the observatory design, which presents the optimization work done to date and outlines the work in progress. 

\subsubsection*{Observatory Requirements}

\begin{figure}[!b]
\begin{minipage}{0.45\textwidth}
		\centering
		\includegraphics[width=\linewidth,trim={0.5cm 0.8cm 0.5cm 0.5cm},clip]{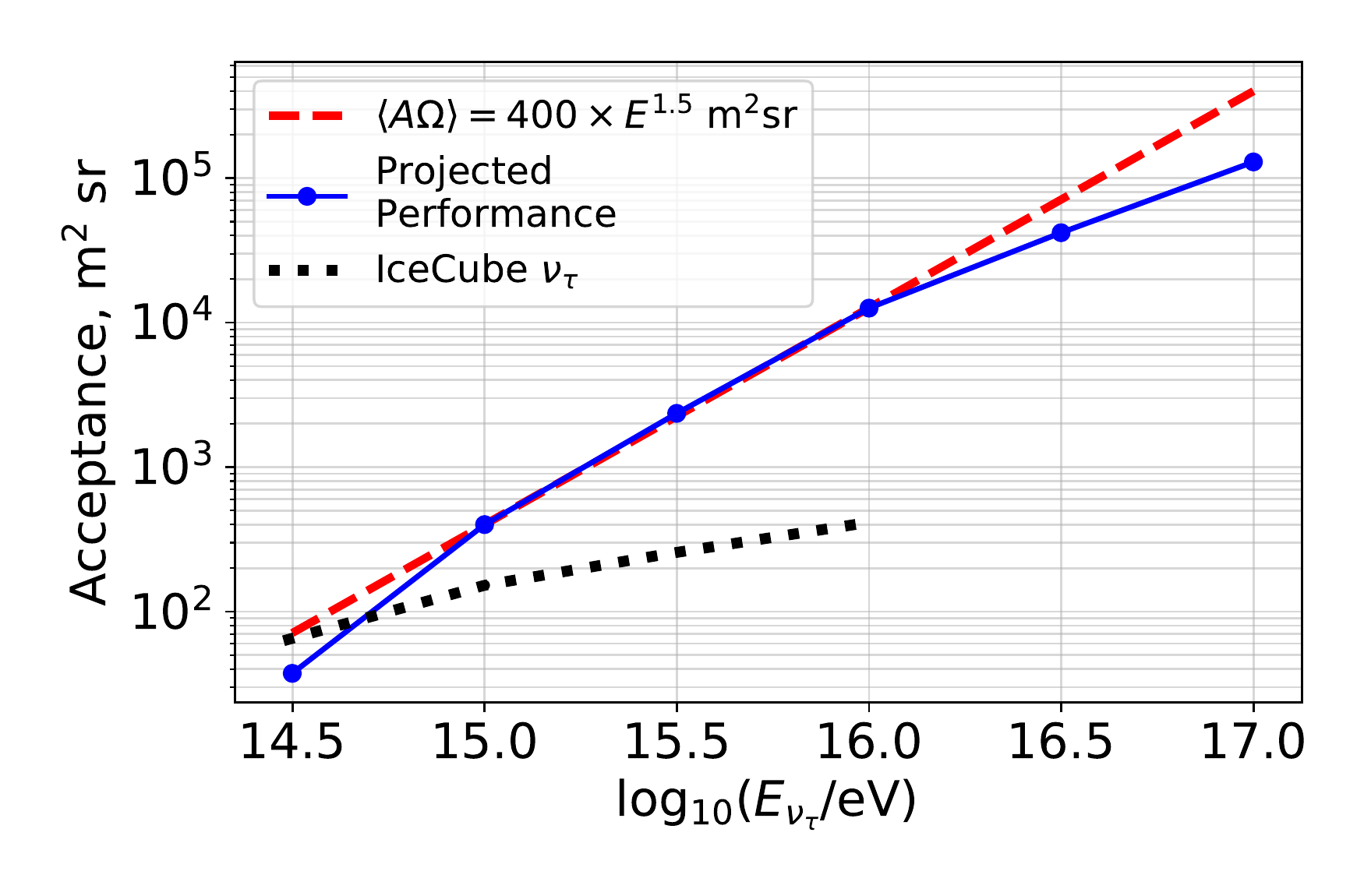}
\end{minipage}
\hspace{0.5cm}
	\begin{minipage}{0.5\textwidth}
	\centering
		\includegraphics[width=\linewidth]{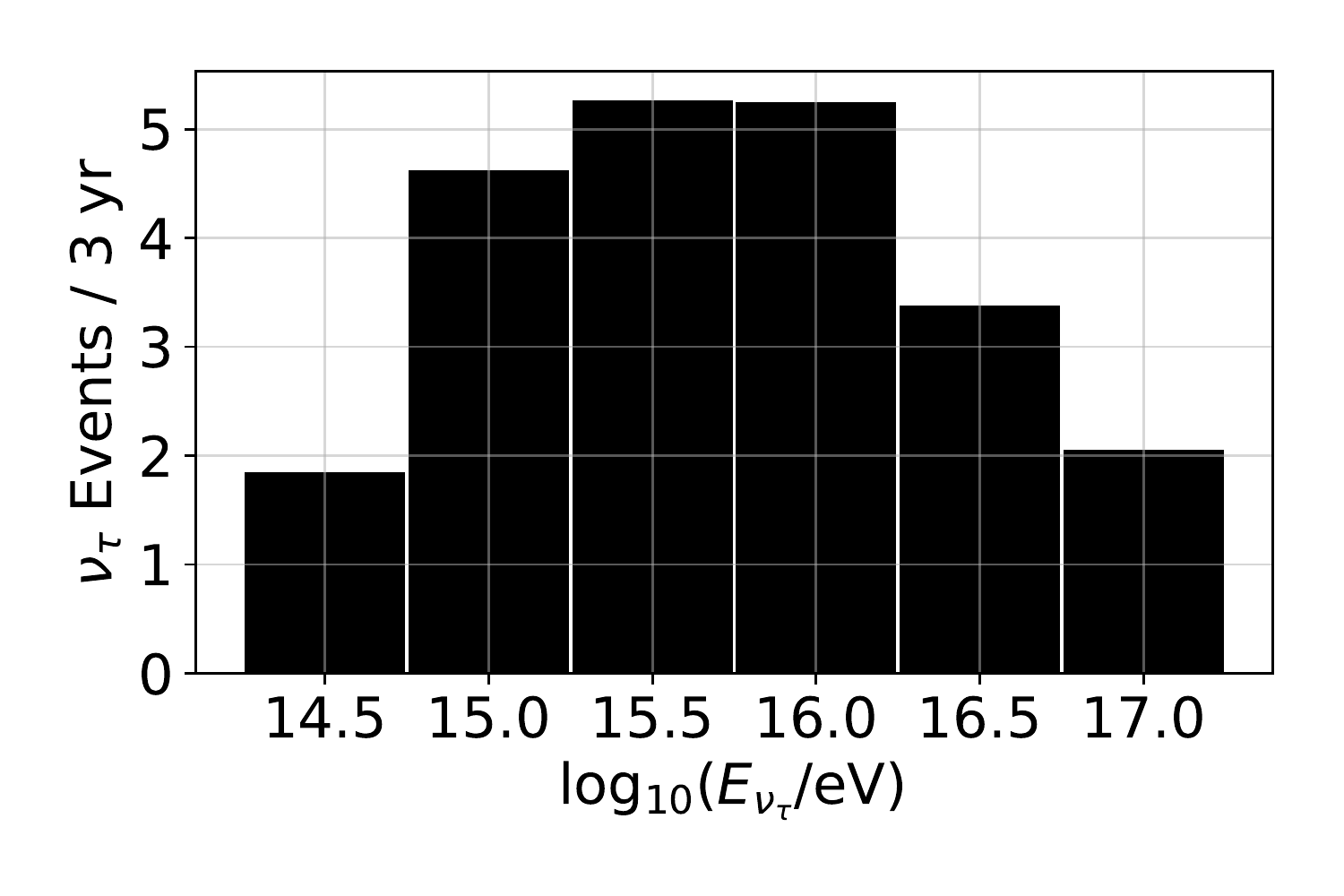}
\end{minipage}
\vspace{-0.5 cm}
\caption{Left: projected performance for the acceptance as a function of energy of an array of m$^3$ water-Cherenkov detectors with $150$~m separation and 22,000 detectors. The reduced acceptance at lower energies ($<$1~PeV) is due to the air showers being too small to trigger the array on the other side of the valley while at high energies ($>$10~PeV) the loss in sensitivity is due to the $\tau$ range exceeding the dimensions of the valley. The $\nu_\tau$ IceCube acceptance~\cite{IC_2013} is shown for comparison. Right: the total expected rate with this acceptance on an extrapolated IceCube flux is 21 events per 3 years of operation with a peak at 3 PeV.}
\label{FIG:acceptance}

\end{figure}

{\bf Diffuse flux acceptance: } The size and number of detectors on this observatory has been designed around the requirement that $\langle A\Omega \rangle\geq 400 \ \mathrm{m^2 \ sr} \ \left(\frac{E}{\mathrm{PeV}}\right)^{1.5}$ at energies between 1-10 PeV, for objective O2, and that the acceptance be at least an order of magnitude greater than IceCube in the 10-100~PeV energy band for objective O1. The expected performance, based on simulations described in the next subsection, is shown in Figure~\ref{FIG:acceptance}. At $\sim\,$1~PeV energy, the IceCube acceptance for $\nu_\tau$ events is $\langle A\Omega \rangle\simeq130$~m$^2$sr~\cite{IC_2013}. Since the detector presented here would be primarily sensitive to $\nu_\tau$ events and not other flavors, we target a $\nu_\tau$ acceptance of 400~m$^2$~sr, which is approximately three times the single-flavor acceptance of IceCube at 1~PeV. The IceCube single power-law flux for cascade events is $\propto E^{-2.5}$~\cite{IceCube_2015, Niederhausen_2017}, which drives the requirement that the acceptance be proportional to $E^{1.5}$. 

The probability of $\tau$ production from $\nu_\tau$ interactions in rock that exit to the atmosphere provides the desired $E^{1.5}$ dependence on the acceptance~\cite{ Dutta_2002, Tseng_2003, Bottai_2003, Yoshida_2004, Jeong_2017, Alvarez-Muniz_2018}. The decay of $\tau$ particles exiting the rock will produce air showers which can be detected with conventional particle counters~\cite{Fargion_1999, Letessier-Selvon_2001}. This technique preferentially detects $\nu_\tau$ events due to the combination of the decay lifetime and energy loss of the $\tau$ produced in a charged-current interaction. Electrons produced by $\nu_e$ events interacting in rock lose energy almost immediately. Muons produced by $\nu_\mu$ events can exit into the atmosphere, but their detection is suppressed by the long muon lifetime and the low probability of interactions that result in significant energy transfers in the atmosphere.

{\bf Energy Resolution: } A significant part of the energy resolution error budget is dominated by the neutrino interaction and $\tau$-decay process. The average energy transfer to the $\tau$ in a charged-current interaction that exits the rock is $\sim\,$80\%. Using the results of~\cite{Alvarez-Muniz_2018}, the exiting $\tau$ has an energy spread of $\sim\,$14\% around the most probable energy. For the purposes of this discussion, uncertainties and resolution values are 1$\sigma$ (68\% confidence) unless otherwise stated. 

The fraction of energy transferred to particles that produce showers (electrons and hadrons) from a $\tau$ decay has an average value of 55\% with an uncertainty of 60\%, based on Pythia simulations~\cite{Pythia} of $\tau$ decays at relativistic energies. 
The $\tau$ decays that produce hadrons are dominant (64.8\%) and the next dominant mode is a decay that produces an electron (17.85\%). TAMBO is not designed to distinguish between electromagnetic and hadronic air showers, although this could be a considered as a future upgrade by adding the capability to estimate the muon content of the air shower.  
The remaining decay modes produce a muon, which we consider undetectable. 

The combined uncertainties of the energy transferred from the neutrino to the $\tau$ and the $\tau$ decay summed in quadrature is $\sim\,$62\%. If the shower energy resolution is $\Delta E/E\leq80\%$, then this would achieve the target $\nu_\tau$  energy resolution $\Delta E/E\leq100\%$ (also summed in quadrature). Air-shower arrays typically perform better than 40\% energy resolution. For example, the CASA-MIA air-shower array achieved a 25\% energy resolution at their threshold energy of 100 TeV and significantly better at higher energies~\cite{Glasmacher_1999}. The Pierre Auger Observatory achieves $<20$\% energy resolution at threshold~\cite{Fenu_2017}. However, these observatories can assume a constraint on where the first interaction of a cosmic ray occurs in the atmosphere. For $\tau$ decays in a deep valley, the position of a $\tau$ decay is not nearly as well constrained since it depends on the lifetime of the $\tau$. This is a range of 50~m at 1~PeV and 5~km at 100~PeV. Initial results, using an estimate of the total estimated number of particles in tanks with signal equivalent to 3 or more particles, constrain the shower energy to within an order of magnitude but with long tails in the distribution. The estimated uncertainties provided are only for mono-energetic neutrinos and the ability to reconstruct energy given a $\nu_\tau$  spectrum is pending. Note, however, that the expected event rate as a function of energy is relatively flat (right panel of Figure~\ref{FIG:acceptance}) which reduces the uncertainties due to spillover in spectral bins. Systematic uncertainties in energy have not been included in this discussion. A study aimed at improving the shower energy resolution using fits to the shower profiles and timing between signals sampled with the detector array will be the subject of future work.  

   
{\bf Pointing Resolution:} The pointing resolution requirement of $\sigma_\theta\leq 1^\circ$ was identified as sufficient to enable astronomical observations in the Astro 2020 science white paper on neutrinos~\cite{Vieregg_2019}. Air-shower arrays typically achieve $\sim$1$^\circ$ pointing resolution~\cite{SPASE_1991, Borione_1994, Auger_2009, Tkachev_2011, IceTop_2013}. This is enabled by the timing resolution of each detector and their separation $D$ via the relation $c \ \sigma_\tau \leq \sqrt{2} \ D \ \cos\theta \ \sigma_\theta$, where the $\sqrt{2}$ factor comes from the fact that at least 2 pairs of detectors are involved in a triggered event. For D=150~m (see Observatory Design section below) and a mountain slope of $35^\circ$, the timing uncertainty required is $\sigma_\tau \leq 7$~ns in the horizontal direction where the acceptance peaks. Photomultiplier tubes (PMTs) with $\sim\,$ns time resolution are common and it is expected that this requirement could be satisfied.

{\bf Flavor Identification:} This array has been designed to be primarily sensitive to $\nu_\tau$ events by virtue of $\nu_\tau$ interactions and $\tau$ propagation in rock. However, there is the potential of having $\lesssim\,$1 background event for every 20 $\nu_\tau$ events (see Backgrounds section below). This tolerance still satisfies the $5\sigma$ sensitivity required by objective O1.

\begin{figure*}[!b]
		\includegraphics[width=0.329\linewidth]{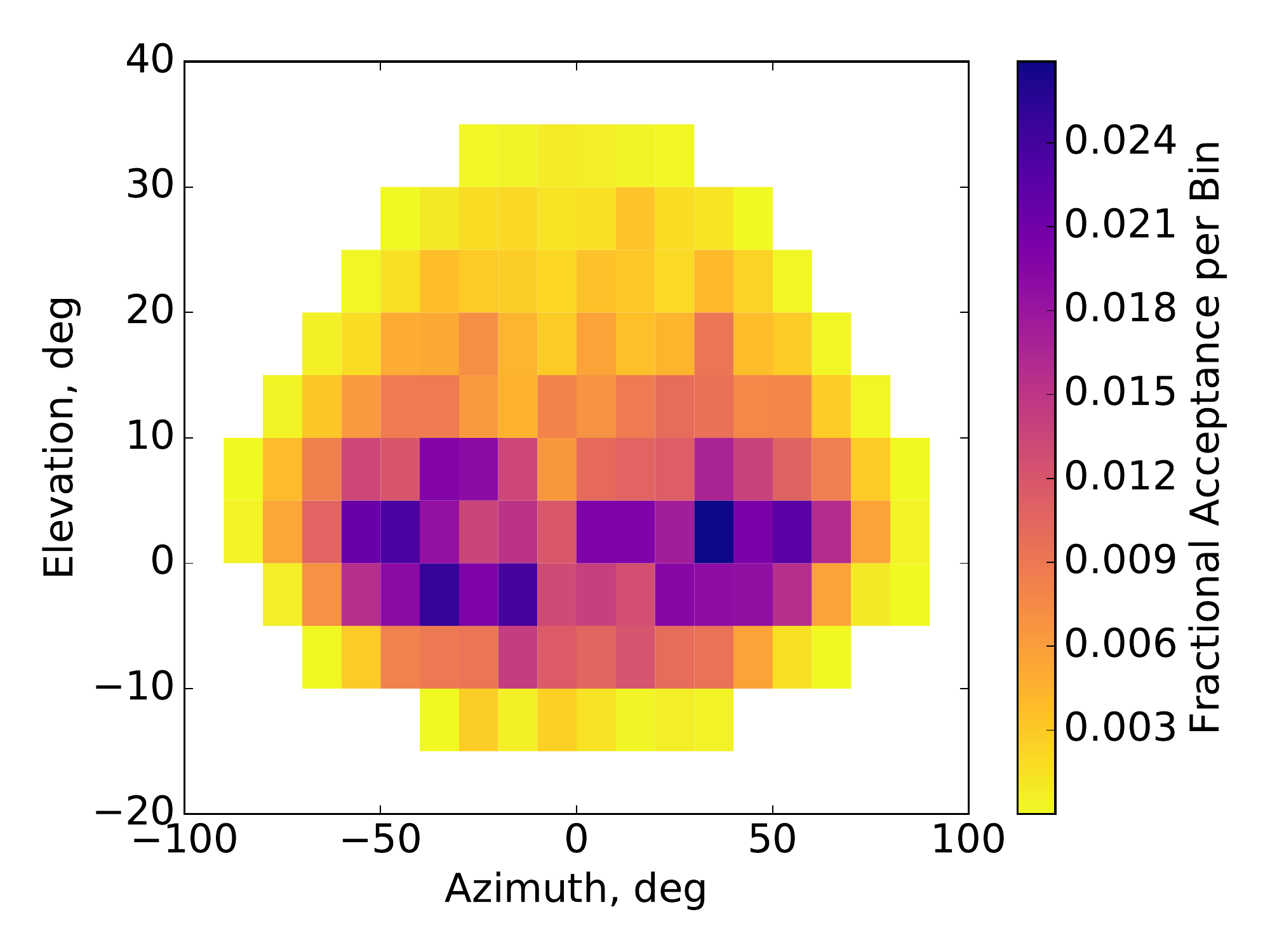}
		\includegraphics[width=0.329\linewidth]{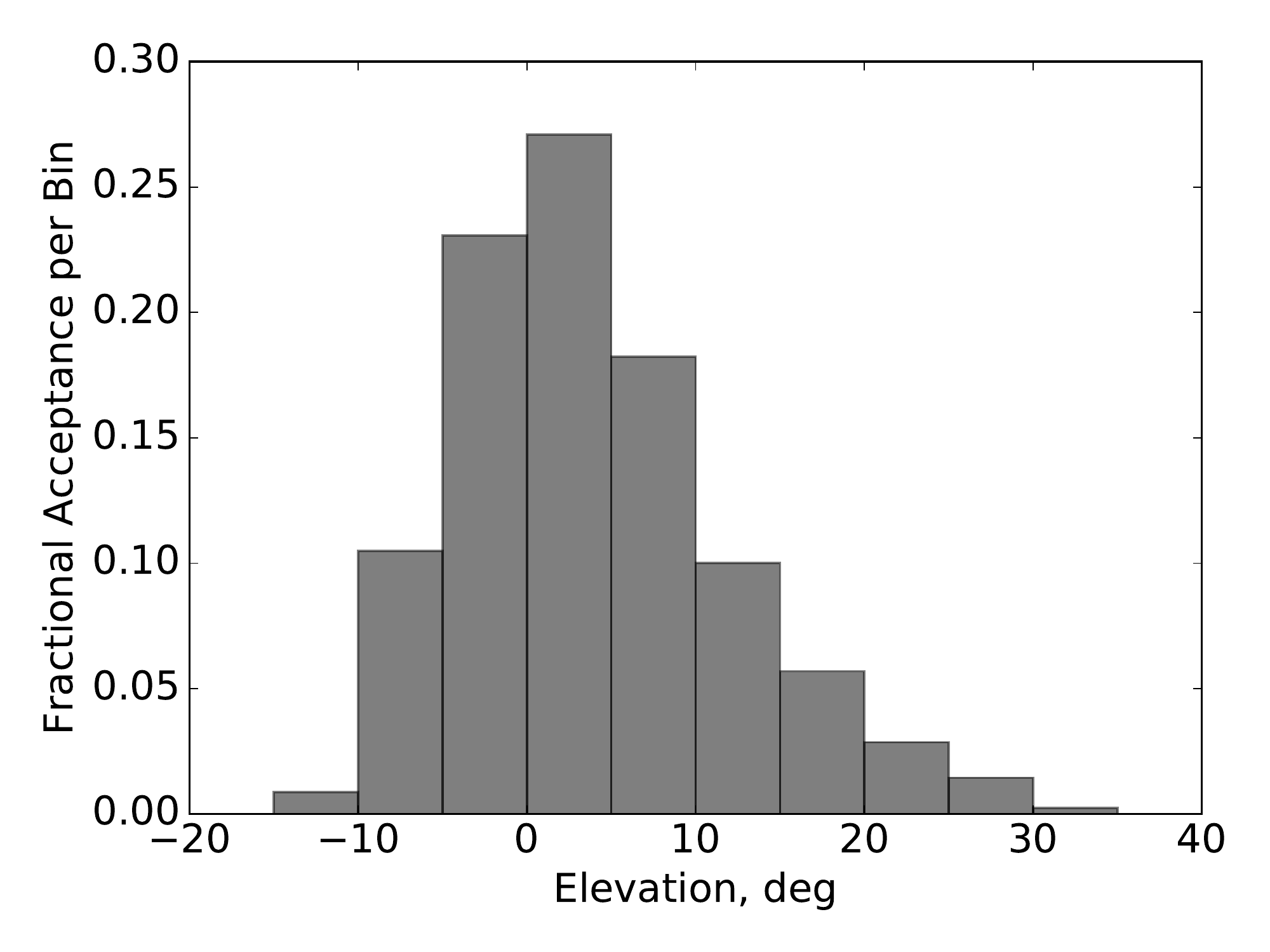}
		\includegraphics[width=0.329\linewidth]{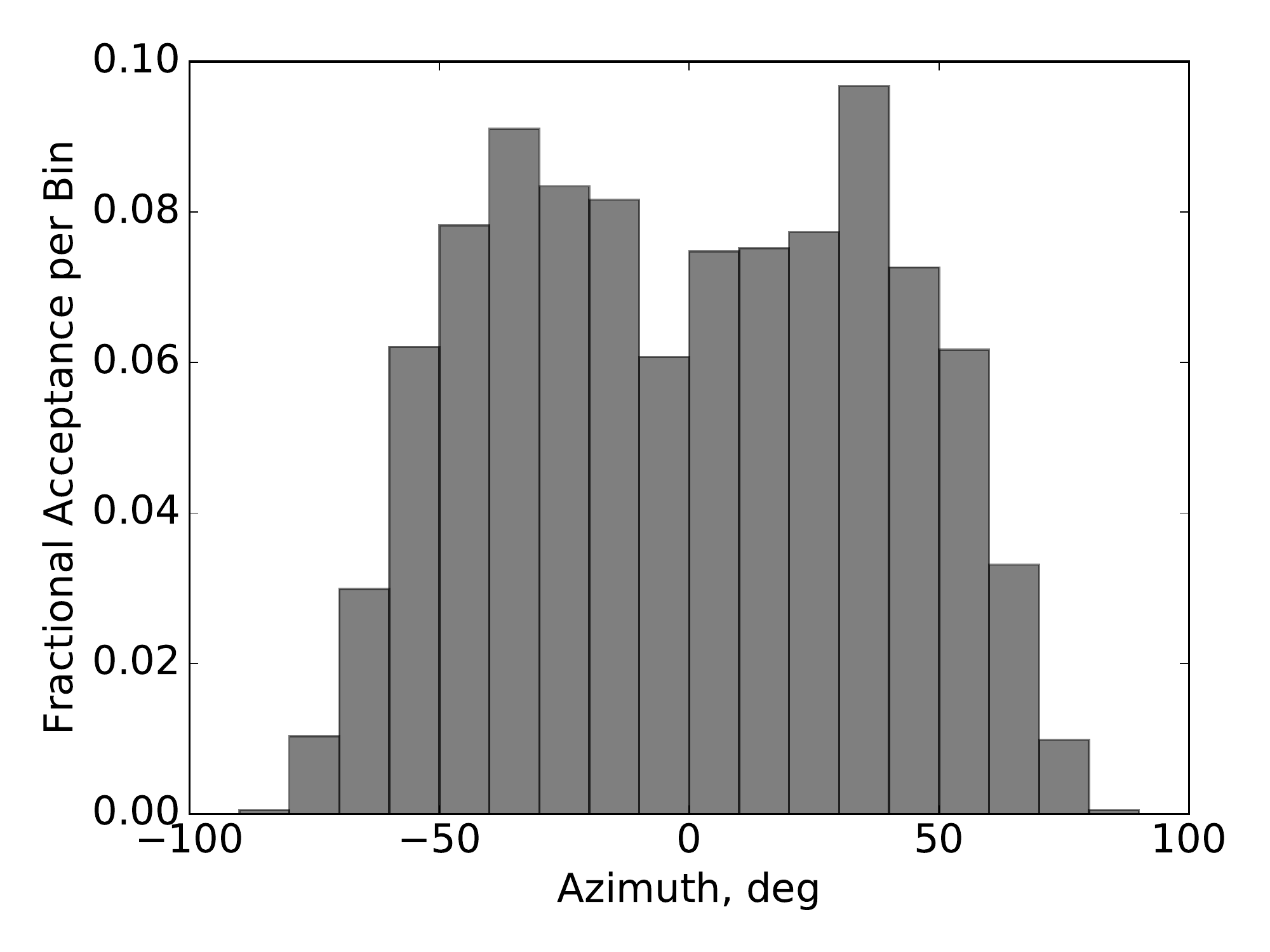}
\caption{Left: the fractional acceptance (normalized to unity) as a function of particle arrival directions that trigger the array. Middle: the elevation angle projection from the figure on the left. Right: the azimuth angle projection from the figure on the left. The majority of particles arrive near the horizontal direction. The span in azimuth is limited by the decay distance and detector density. }
\label{FIG:arrival_directions}
\end{figure*}

{\bf Point Source Effective Area:} While this array was designed by optimizing for the diffuse flux acceptance, it is expected to have a sizable point source effective area. The field of view is restricted to the range of elevation angles subtended by the mountain facing the detector array ($\sim\,$25$^\circ$) and azimuthally by the air-shower longitudinal profile ($\sim\,$10~km) corresponding to $\pm60^\circ$ (see Figure~\ref{FIG:arrival_directions}). 
Based on the behavior of the diffuse acceptance flux, we have a goal of achieving a point source effective area of ${\langle A\rangle\gtrsim300 \ \mathrm{m}^2\times(E/\mathrm{PeV})^{1.5}}$.  This would roughly match the IceCube effective area of $\sim\,$200~m$^2$ at 1 PeV~\cite{IceCube_2017_pt_src}, and exceed it by more than an order of magnitude at 100 PeV. 

{\bf Sky Coverage:} The sky coverage can be optimized to cover a wide region, by spreading the detectors across various valley orientations and locations across the world, or over a narrow region optimizing for point source sensitivity, by clustering the detectors over a long valley that is roughly oriented towards the same direction in the sky. We plan to optimize for the latter, within the geometrical constraints of existing valleys, since it is more relevant to multi-messenger astronomy. The Colca Valley site (Peru) used for our current study has regions that would point North/South as well as East/West. With the $\gtrsim\,$2~km altitudes over which detectors are populated, it is expected that the detector would cover $>\,$0.1~sr instantaneous field of view (see Figure~\ref{FIG:arrival_directions}). With the rotation of the Earth, the integrated coverage is expected to cover $>\,$0.5~sr. It is possible to consider more than one site and detector layout to increase sky coverage or point to specific sources of interest. These trades will be considered in future stages of the observatory design layout.

{\bf Backgrounds:} In order to achieve the $5\sigma$ sensitivity required by objective 1, the backgrounds need to be suppressed to $\lesssim 1$ event per 3 years, based on the expectation of $\sim\,$20 events per 3 years. We have considered four main sources of background. 

The first source of background considered is due to random coincidences of small showers and vertical muons mimicking a $\tau$ air shower coming from the opposite side of the valley. The rate of small showers and muons for an array of the size proposed here is in the scale of kHz. Strategies to minimize this background are under investigation. The current trigger design requires that coincident triggering of at least three neighboring tanks, each with the equivalent signal of 3 particles each as well as the summed signal of triggered tanks be greater than the equivalent of 30 particles. The summed signal requirement is motivated by there typically being one tank closest to the shower core with high signal and is meant to reduce backgrounds from coincident small showers. Additional discrimination strategies include the use of shower profile fits and signal timing constraints.

The second source of background considered is the directional reconstruction uncertainty of cosmic ray air showers near the sky-mountain boundary that would appear to point to the mountain facing the detector. With $\sigma_\theta = 1^\circ$ resolution, this background is efficiently suppressed to $<0.1$ events per 3 years by setting a cut on events that have a direction more than 
$8^\circ$ below the sky-mountain boundary.  Figure~\ref{FIG:arrival_directions} shows the fractional acceptance at 1~PeV as a function of direction. The sky-mountain boundary is at an elevation angle of $\sim\,$35$^\circ$ and cutting out $8^\circ$ below it is of minimum impact to the acceptance. 

The third source of background considered is the high-energy muons produced by cosmic-ray air showers~\cite{Enberg_2008} behind the mountain that would propagate through to produce an air shower in the valley. If a muon interacts with air via nuclear bremsstrahlung, a sizable energy transfer to a photon producing an electromagnetic air shower may occur (although the probability is estimated to be $<0.4\%$ based on AIRES simulations~\cite{AIRES}). The energy loss of muons in rock combined with the expected spectrum of high-energy muons results in this background being suppressed below the detector threshold for trajectories of $>4$~km of rock. The regions with $<4$~km of path length through rock are already cut out by the cosmic-ray air shower cut described in the paragraph above. 

The final source of background considered is from prompt $\nu_\mu$ produced in air showers~\cite{Enberg_2008}. Muon neutrinos propagating in rock can undergo a charged-current interaction that produces a muon. The range of the muon is in the scale of kilometers meaning that the production of muons per neutrino exiting the rock into the atmosphere is greater than that of $\tau$ particles. However, the muon lifetime is much longer and it is highly unlikely to decay before reaching the detector. There is still a small probability ($<0.4$\%) that an energetic muon can transfer a significant fraction of its energy to a photon via a nuclear bremsstrahlung interaction resulting in an electromagnetic air shower. Preliminary estimates result in an expected value of 1 muon neutrino background event per 20 $\nu_\tau$ events for energies $\geq 1$~PeV. More accurate bounds, which are expected to be lower, are currently being studied with more detailed simulations.

\subsubsection*{Observatory Design}
The approach to observatory design was to identify a deep valley first and then optimize the array design to achieve the required diffuse flux acceptance while minimizing the number of stations. 

{\bf Deep Valley Sites:} The deep valleys applicable for this concept need to have a width of 3-10~km, driven by the longitudinal profile of the electromagnetic component of air showers, and km-scale depth for efficient use of a detector array. There also needs to be $\sim\,$100~km length of mountain face (both sides can be used) to achieve the required acceptance. Since TAMBO relies on the use of water-Cherenkov detectors, it is preferable to have a source of water nearby with existing infrastructure to reduce the cost of construction. 
Although this list is by no means exhaustive, we have considered the following four of the world's deepest valleys. These are Hells Canyon in the United States of America, Yarlung Tsangpo Grand Canyon in China, Cotahuasi Canyon and the Colca Canyon, both in Arequipa, Peru. 

\begin{figure}[!t]
	\centering
		\includegraphics[width=0.6\linewidth]{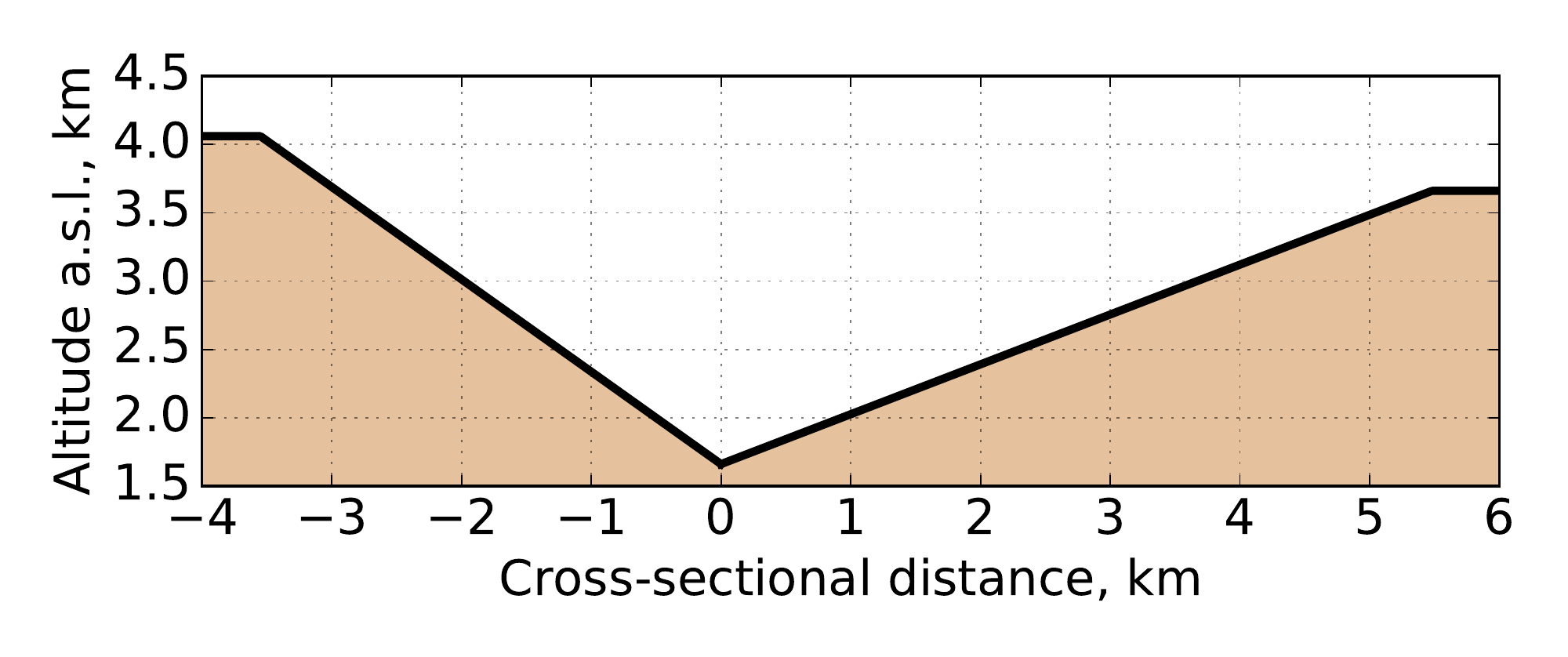}
	\caption{A cross section of the Colca Valley near 15$^\circ$38'35"~S 72$^\circ$08'55"~W. The distance between the mountain faces is $\sim\,$4.5~km at half the valley depth. For the simulation results presented, the left side of the valley was populated with a detector array. 
	}
	\vspace{-0.25cm}
	\label{FIG:Colca_cross_section}
\end{figure}

The Colca Canyon has a river running at its base, road access, and nearby towns. These features make it an attractive site. The Colca Canyon has regions with up to 3,270~m depth with $\gtrsim$100~km length of mountain face. In Figure~\ref{FIG:Colca_cross_section} we show an approximate cross section of the canyon located near 15$^\circ$38'35"~S 72$^\circ$08'55"~W that we used as the fiducial geometry for the results in this white paper. A more detailed optimization study using the topography along the valley will be the subject of a future study.

{\bf Detector Array Design:} Given that a large area has to be covered to reach the required acceptance, we have taken into consideration minimizing the size and number of detectors needed. The first trade was between scintillator detectors and water-Cherenkov counters. The latter was chosen because for every charged particle in an air shower, there are roughly ten times the number of high-energy photons. While scintillators can be made sensitive to photons by adding lead plates, the approach is not as efficient as a water-Cherenkov tank.

\begin{figure}[!t]
\begin{minipage}{0.5\textwidth}
	\centering
		\includegraphics[width=\linewidth]{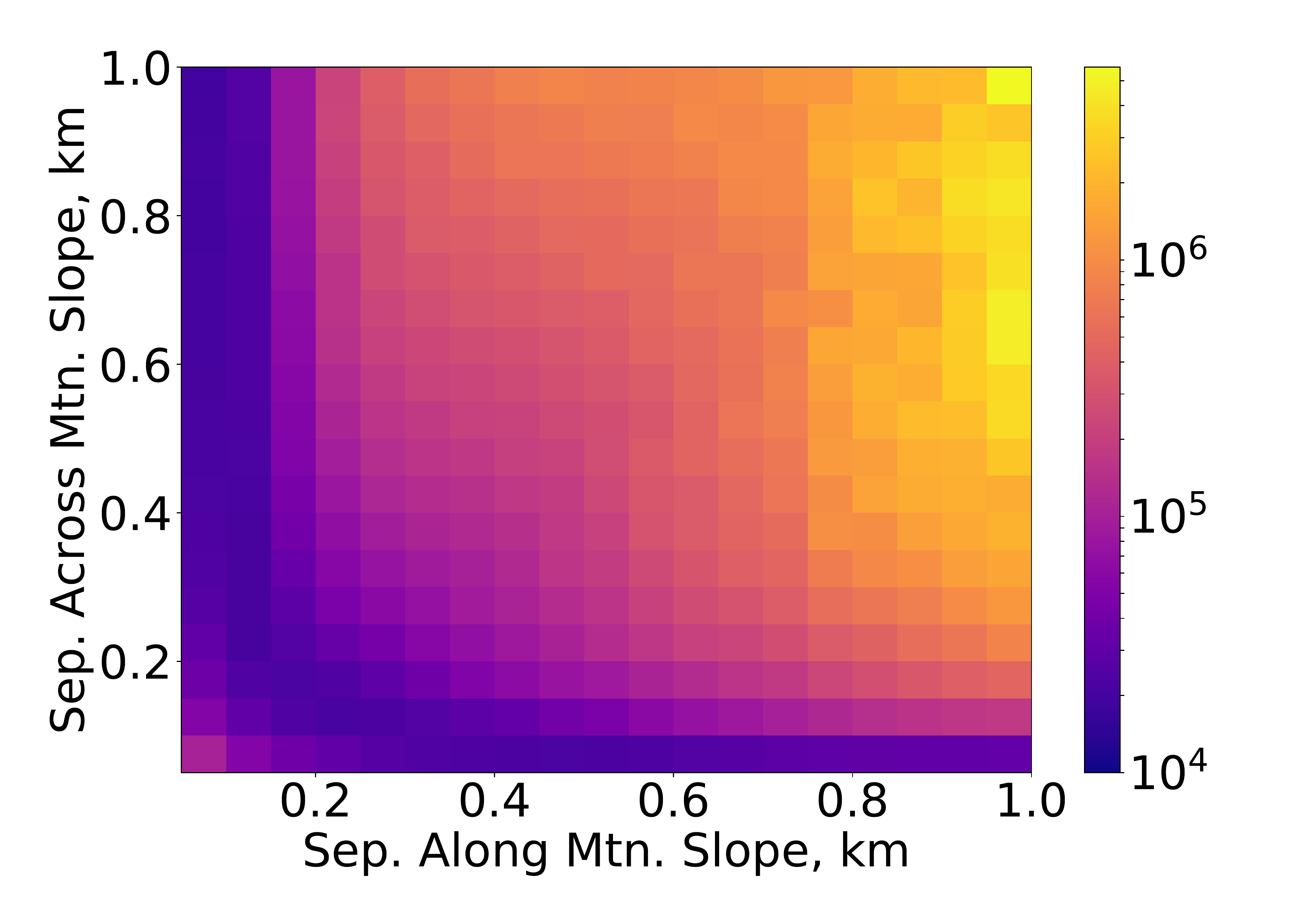}
\end{minipage}
\begin{minipage}{0.5\textwidth}
	\centering
		\includegraphics[width=\linewidth]{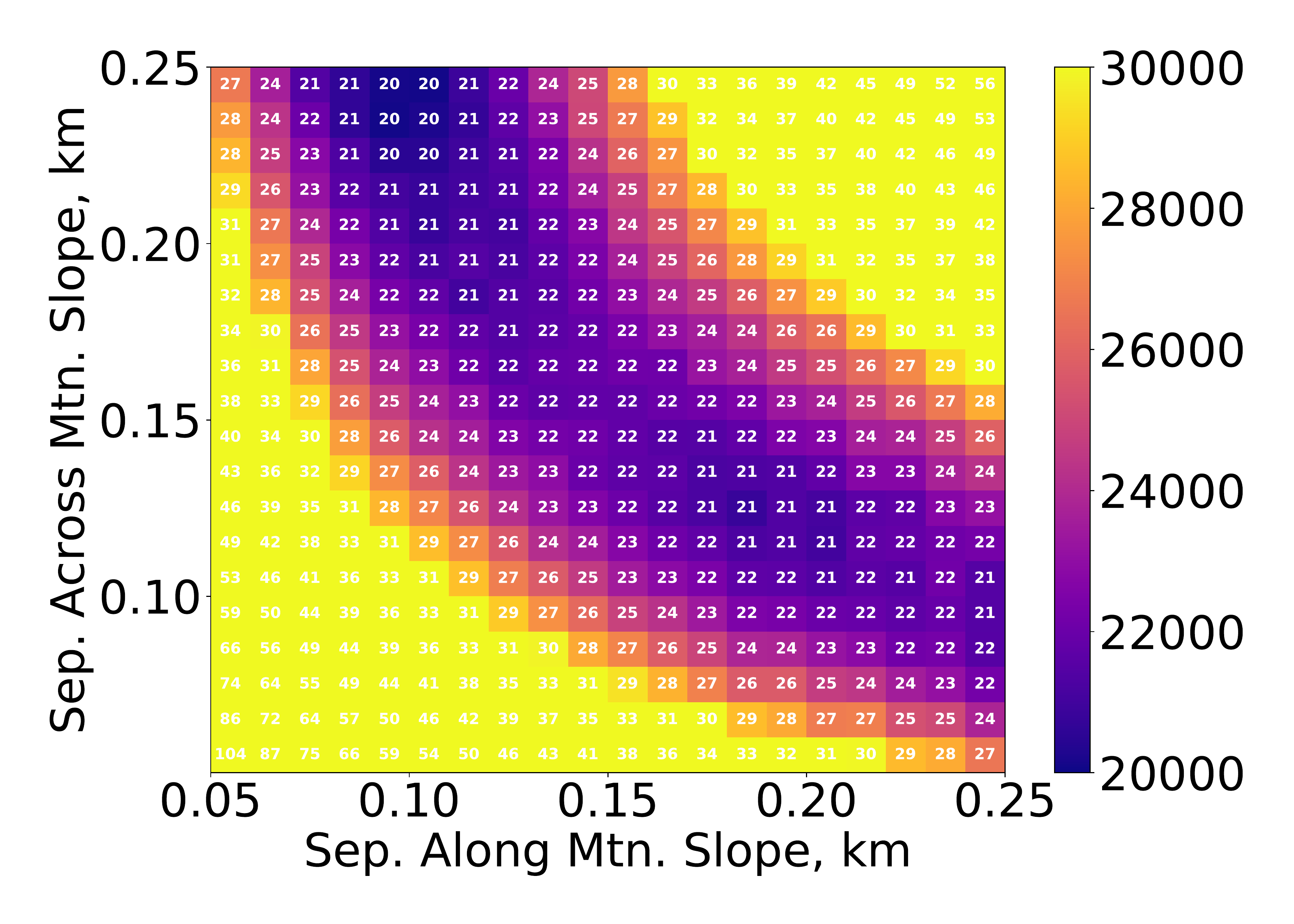}
	\end{minipage}
	\caption{Left: broad search for the minimum number of stations, shown in the color scale, as a function of separation along and across the mountain slope, required to obtain $\langle A\Omega \rangle$=400~m$^2$sr at 1 PeV. The region of interest is where the separations in each direction are comparable.
	Right: Refined search for the minimum number of stations near the minimum region on the left plot. Note the change from logarithmic scale in the left figure to linear scale in the right. The number of stations, corresponding to the color scale, is displayed in white text in each bin. There is a trough of values that achieve the target acceptance with $\sim\,$22,000 stations (with statistical fluctuations). The optimal separation is $\sim\,$150~m in both directions.}
	\label{FIG:optimize}
\end{figure}

For the trigger design, we started with the approach of the Pierre Auger Observatory of requiring three tanks with a time over threshold trigger requiring the equivalent of at least three particles entering each tank~\cite{Auger_Trigger}. For the trigger, we add the requirements that the sum of signals in tanks exceeding the three-particle equivalent is at least the equivalent of 30 particles to reduce backgrounds from random coincidence. This is based on the expectation that tanks near the air-shower core will have a significant signal. 

Based on the density of photons and charged particles in the shower, a water-Cherenkov detector with cross-sectional area of $\sim\,$m$^2$ and $\sim\,$m$^3$ volume is sufficient to trigger efficiently. We ran a simulation to determine the minimum number of tanks needed (see Figure~\ref{FIG:optimize}) and arrived at 22,000 of them with 150~m spacing in a triangular grid on the mountain slope. There is a trade between the number of tanks and their size. However, since these detectors have to be placed on steep mountain slopes (up to $35^\circ$) it is important to keep the size of the tanks small to facilitate deployment. 

The current concept for the array has the water tanks deployed as lines that run along the slope of the mountain. The power and data wire connections run along the line to an electronics box at the base or top of the mountain (whichever is more accessible). The electronics boxes for each line will be separated by 150~m which allows for wired connections between adjacent boxes to run the trigger logic. The goal is to provide sufficient real-time identification of the cosmic-ray air shower background so that the data of interest can be transferred via a modest-rate data connection and be used for multi-messenger alerts. 

One advantage of TAMBO is that it does not require any new technologies. Water tank detectors with PMTs have been in operation since the 1960's. PMTs with nanosecond time resolution are commercially available. For the design of water tanks there is a trade to be done between reflective and non-reflective inner surfaces that affects the photon detection efficiency and timing of air showers. However, these tools have been developed for HAWC and SWGO where we have overlapping collaborators. The electronics used for the detector are expected to be relatively simple. The systems would be connected by lines that go along the mountain slope and connections between neighboring lines would be done at the base or peak of the mountain. Standard trigger and veto techniques would be applied to reduce the data rates to manageable levels. The Colca Canyon has a major advantage in that it is populated and road access to large portions of the valley are available. 

\subsection*{List of the interested scientists in the community}

The list of scientists actively involved in TAMBO is provided in the cover page. The author list of~\cite{Vieregg_2019} and~\cite{Ackerman_2019} represents the broader neutrino astrophysics community interested in this detector as it is designed in direct response to the goals set for the next decade. The study is in the early phases of development and the team is expected to grow as the concept finalizes and prototype demonstrations become available. A detector of this magnitude is expected to have international collaborations and contributions from multiple funding sources as with the Pierre Auger Observatory and IceCube. The core team developing this study have overlapping and complementary roles in Auger, ANITA, HAWC, and SWGO, where hardware, operations, and analysis expertise is highly relevant. 

The Colca Canyon is currently an attractive site for developing and implementing this concept. We have partnered with several Peruvian institutions that have shown overwhelming support for development of large-scale astrophysical observatories. We are currently partnering with the {\it Universidad Nacional San Agust\'in} (UNSA) of Arequipa in Peru as they are a large university that is close to the site and they are also becoming involved in the development of SWGO in Peru. UNSA will also be involved in the academic aspects with participation from professors and research students. They also have significant competed funds available for pure and applied research that can be used to develop a neutrino observatory in the region of Arequipa. The {\it Consejo Nacional de Ciencia Tecnolog\'ia e Innovaci\'on Tecnol\'ogica} (CONCYTEC), which facilitates the interactions between government and research institutions, has also taken an active interest in promoting these developments. In addition, we have partnered with Alberto Gago and Jose Bazo of the {\it Pontifica Universidad Cat\'olica del Peru}. They are currently active collaborators in several international neutrino observatories and experiments such as IceCube and DUNE. 

\subsection*{Current status and expected challenges}

TAMBO is currently in the concept development phase. We have identified the detector requirements to obtain the targeted $\nu_\tau$ sensitivity. As mentioned in the methodology section, we are refining our background modeling to ensure adequate suppression. These backgrounds are common to air shower and neutrino detectors and the techniques to model and suppress them exist. Drs. Jose Bellido and Andres Romero-Wolf have recently partnered with Professor Rolando Perca and students at the {\it Universidad Nacional San Agust\'in} (UNSA) of Arequipa with the goal of prototyping detectors in Arequipa and experimentally determine the feasibility of TAMBO. As we mentioned in the Methodology section, TAMBO does not require any new technologies and no challenges are expected on that front. However, TAMBO does require that the detector be deployed in the steep mountain face of the Colca Valley, which may be a challenge and requires further study.

\subsection*{Timeline}

We present a notional schedule of how the array could be implemented over the span of the next decade. It's worth noting that IceCube was deployed over a period of $\sim\,$6~years with constraints of South Pole seasons, while Auger was deployed over the time period of $\sim\,$4 years. Although TAMBO has about an order of magnitude more detectors, each of them is simpler and smaller than a Pierre Auger Observatory water-Cherenkov tank, and the deployment does not require drilling as in the case of IceCube. 

It is expected this team will need another year to fully address the questions of background suppression and finalize the initial array design. After publication of the detailed concept study in 2020, prototype development would begin. Our collaboration has overlap with members of the Pierre Auger Collaboration as well as SWGO, which is developing water tank detectors and also currently considering Peru as a site for their observatory. Once the initial prototype, consisting of a few water tanks, is built and tested to validate hardware performance in 2021, there would be a broader call for collaborators to develop and operate the observatory. At this time, the development of a field-deployable prototype would begin with the goal of installing two parallel lines connected to each other at a deep-valley site in 2023 to operate for a year. This prototype would validate on-site performance requirements, calibration, and background rejection by 2024. In 2025, the development of a 200 water tank array would begin with construction in 2026 and operations in 2027. Similarly, the 2,000 tank array will be built and operated in 2028-2029. Note that this array would be expected to detect $\sim\,0.7$~events per year thereby producing the first detection or best limits at energies $\gtrsim$10~PeV. During this development, the mass production and deployment of detectors would be one of driving goals. The construction of the 22,000 tank array would begin in 2029 with deployment and operations occurring through 2030.

\subsection*{Construction and operational costs}

At this early stage in the development it is not possible to provide accurate cost estimates. It is expected that this ground-based observatory would be in the $\sim\$100$M category. The construction costs of some analogous detectors are the Pierre Auger Observatory, which was in the \$50M category. The deep-valley detectors are smaller but there are more of them and the costs of deploying lines on steep mountain faces has to be better understood through prototype development. The IceCube construction costs were in the \$100M category but TAMBO does not require drilling, and access to the Colca Valley is significantly easier thanks to existing infrastructure. The best analogue for operation costs is the Pierre Auger Observatory, which is $\sim$\$2M per year.
 
\vspace{0.25 cm}

\noindent{\it Cost disclaimer:} The cost information contained in this document is of a budgetary and planning nature and is intended for informational purposes only. It does not constitute a commitment on the part of JPL and/or Caltech.

\subsection*{Computing requirements}
The data obtained will be pulses measured with photomultiplier tubes with the array triggering at ~kHz rate due to the background of cosmic-ray air showers. These are expected to be readily filtered by directional reconstruction based on the timing of the signals between different tanks. This level of filtering can be done efficiently in real time. This portion of the computing would be built into the real time operations of the array. For the data analysis, the computing requirements are not expected to drive the array. This is still under investigation as it depends on the array's ability to efficiently flag backgrounds in real time vs in post-processing. In analogy with HAWC, scaled for event rate and size of detector, it is estimated that <500 cores will suffice. This is not expected to be a cost driver.
 
\vspace{0.25 cm}

\noindent{\it Acknowledgements:} Part of this work was carried out at the Jet Propulsion Laboratory, California Institute of Technology, under a contract with the National Aeronautics and Space Administration. A. R.-W. thanks the Caltech/JPL President's and Director's Research and Development Fund. Copyright 2019. All rights reserved. 
 


\pagebreak


\end{document}